\documentclass[11pt,dvips]{article}

\usepackage{epsfig,times} 
\usepackage{picinpar}
\makeatletter
\renewcommand{\section}{\@startsection{section}{1}{0in}
	{0.4\baselineskip}{0.1\baselineskip}{\Large\bf}}
\renewcommand{\subsection}{\@startsection{subsection}{2}{0in}
	{0.25\baselineskip}{-\baselineskip}{\large\bf}}
\renewcommand{\subsubsection}{\@startsection{subsubsection}{3}{0in}
	{0.1\baselineskip}{-\baselineskip}{\normalsize\bf}}
\makeatother
\def\fdg{\hbox{$.\!\!^\circ$}}
\def\degr{^\circ}

\begin{document}

%
\makeatletter\newcommand{\ps@icrc}{
\renewcommand{\@oddhead}{\slshape{OG.2.2.16}\hfil}}
\makeatother\thispagestyle{icrc}
%
\markright{OG 2.2.16}
\begin{center}
%
{\LARGE \bf Search for TeV Gamma-Rays from Shell-Type \\Supernova Remnants}
\end{center}

\begin{center}
%
%
{\bf R.W. Lessard$^{1}$, I.H. Bond$^{2}$, P.J. Boyle$^{3}$, 
S.M. Bradbury$^{2}$, J.H. Buckley$^{4}$, A.M. Burdett$^{2,5}$, 
D.A. Carter-Lewis$^{6}$, M. Catanese$^{6}$, M.F. Cawley$^{7}$, 
S. Dunlea$^{3}$, M. D'Vali$^{2}$, D.J. Fegan$^{3}$, S.J. Fegan$^{5}$, 
J.P. Finley$^{1}$, J.A. Gaidos$^{1}$, T.A. Hall$^{1}$, A.M. Hillas$^{2}$, 
D. Horan$^{3}$, J. Knapp$^{2}$, F. Krennrich$^{6}$, S. Le Bohec$^{6}$, 
C. Masterson$^{3}$, J. Quinn$^{3}$, H.J. Rose$^{2}$, F.W. Samuelson$^{6}$, 
G.H. Sembroski$^{1}$, V.V. Vassiliev$^{5}$, T.C. Weekes$^{5}$}\\
{\it
$^1$Department of Physics, Purdue University, West Lafayette, IN 47907, USA\\
$^2$Department of Physics, Leeds University, Leeds, LS2 9JT, UK\\
$^3$Experimental Physics Department, University College, Belfield, Dublin 4,
    Ireland\\
$^4$Department of Physics, Washington University, St. Louis, MO 63130, USA\\
$^5$Fred Lawrence Whipple Observatory, Harvard-Smithsonian CfA, P.O.
    Box 97, Amado, AZ 85645-0097, USA\\
$^6$Department of Physics and Astronomy, Iowa State University, Ames,
    IA 50011-3160, USA\\
$^7$Physics Department, National University of Ireland, Maynooth, 
    County Kildare, Ireland}
\end{center}

\begin{center}
{\large \bf Abstract\\}
\end{center}
\vspace{-0.5ex}
%
%
If cosmic rays with energies $<$100 TeV originate in the galaxy and
are accelerated in shock waves in shell-type supernova remnants (SNRs),
gamma-rays will be produced as the result of proton and electron
interactions with the local interstellar medium, and by inverse
Compton emission from electrons scattering soft photon fields. We
report on observations of two supernova remnants with the Whipple
Observatory's 10m gamma-ray telescope. No significant detections have
been made and upper limits on the $>$500 GeV flux are
reported. Non-thermal X-ray emission detected from one of these
remnants (Cassiopeia A) has been interpreted as synchrotron emission
from electrons in the ambient magnetic fields.  Gamma-ray emission
detected from the Monoceros/Rosette Nebula region has been interpreted
as evidence of cosmic-ray acceleration. We interpret our results in
the context of these observations.
%

\vspace{1ex}

%
%
\section{Introduction:}
\label{section:introduction}
\subsection{Hadronic Gamma-Ray Production:}
\label{section:hadronic}
If SNRs are sites for cosmic-ray production, there will be interactions
between the accelerated particles and the local swept-up interstellar
matter. Drury, Aharonian and Volk (DAV) (1994) and Naito and Takahara
(1994) have calculated the expected gamma-ray flux from secondary pion
production using the model of diffusive shock acceleration. 
Recent observations above 100~MeV by the EGRET instrument on the {\it
Compton Gamma-Ray Observatory} have found gamma-ray signals associated
with at least three SNRs: IC 443 and gamma-Cygni (Esposito et al. 1996)
and the Monoceros SNR - Rosette Nebula region (Jaffe et
al. 1998). These results have been interpreted as evidence for cosmic
ray acceleration using the model of DAV. A search for TeV emission
from IC 433 and gamma-Cygni has already been reported by the Whipple
Collaboration (Buckley et al. 1998). Here we present observations of
the Monoceros SNR - Rosette Nebula region.

\subsection{Leptonic Gamma-Ray Production:}
\label{section:leptonic}
Gamma rays will also be produced as the result of electron
bremsstrahlung and by inverse Compton emission from electrons
scattering soft photon fields. Non-thermal X-ray emission, observed
from several SNRs including SN 1006 (Koyama et al. 1995),
and Cassiopeia~A (Allen et al. 1997), has been
interpreted as synchrotron radiation of high energy electrons in the
ambient magnetic fields.  The recent detection of SN~1006 at TeV
energies by CANGAROO (Tanimori et al. 1998) has been interpreted as
inverse Compton emission from these electrons scattering microwave
background photons as predicted by Pohl (1996), and Mastichiadis and
de Jager (1996). Here we report on observations of Cassiopeia~A.

\section{Observations and Analysis Techniques:}
The observations reported here were taken with the Whipple
Observatory's high energy gamma-ray telescope situated on Mount
Hopkins in southern Arizona (Cawley et al. 1990).  A camera,
consisting of photomultiplier tubes (PMTs) mounted in the focal plane
of the reflector, detects the Cherenkov radiation produced by
gamma-ray and cosmic-ray air showers from which an image of the
Cherenkov light can be reconstructed. For the observations reported
here, the camera consisted of 331~PMTs (each viewing a circular field
of 0\fdg25 radius) with a total field of view of $4.8\degr$ in
diameter.

We characterize each Cherenkov image using a moment analysis (Reynolds
et al. 1993). The roughly elliptical shape of the image is described
by the {\em length} and {\em width} parameters and its location,
orientation and skewness within the field of view are given by the
{\em distance}, $\alpha$ and {\em asymmetry} parameters,
respectively. We also determine the two highest signals recorded by
the PMTs ({\em max1, max2}) and the amount of light in the image ({\em
size}). Gamma-ray events give rise to more compact shower images than
background hadronic showers and are preferentially oriented and skewed
towards the putative source position in the image plane. By making use
of these differences, a gamma-ray signal can be extracted from the
large background of hadronic showers.

\subsection{Centered Point Source Observations:}
The traditional mode of observing potential point sources with the
Whipple Observatory gamma-ray telescope is track the putative source
position continuously for 28 minutes. The standard gamma-ray selection
method utilized by the Whipple Collaboration is the Supercuts98. These
criteria were optimized on contemporaneous Crab Nebula data to give
the best sensitivity to point sources. A combination of Monte Carlo
simulations and results on the Crab Nebula indicate that this analysis
results in an energy threshold of $\sim 500$ GeV and an effective
collection area of $3.0 \times 10^8 {\rm cm}^2$. A detailed
description of the methods used to estimate the energy threshold and
collection area are given elsewhere (Mohanty et al. 1998).

To estimate the expected background, we use those events which pass all
of the Supercuts98 criteria except orientation (characterized by the
$\alpha$ parameter). We use events with $\alpha$ between $20\degr$ and
$65\degr$ as the background region. A detailed description of this
method can be found elsewhere (Catanese et al. 1998). 

\subsection{Extended or Offset Source Observations:}
When the position of the putative source is not at the center of the
field of view, or is not precisely known (e.g., EGRET error circles or
sources of extended emission) a different strategy must be employed.
A detailed description of the technique utilized can be found
elsewhere (Lessard et al. 1997). 

Briefly, the method uses additional features of the images of
gamma-ray induced showers to provide a unique arrival direction on an
event by event basis. First, events are selected as gamma-ray like
based on their compactness. The determination of the arrival direction
of the candidate gamma-ray events is then accomplished by making use
of the orientation, elongation and asymmetry of the image. Monte Carlo
studies have shown that gamma-ray images are a) aligned towards their
source position on the sky b) elongated in proportion to their impact
parameter on the ground and c) have a cometary shape with their light
distribution skewed towards their point of origin in the image
plane. Results on the Crab Nebula indicate that the angular resolution
function for the telescope using this technique is a Gaussian with a
standard deviation of $0.12\degr$ (Lessard et al. 1997). The efficacy
of the method is demonstrated by observations in which the center of
the field of view is offset from the Crab Nebula position.  A
combination of Monte Carlo simulations and results on the Crab Nebula
indicate that this analysis results in an energy threshold of $\sim
500$ GeV and an effective collection area of $3.0 \times 10^8 {\rm
cm}^2$ for a source at the center of the field of view and is reduced
for offset sources.

The analysis of data from a point source, offset from the center of
the field of view, involves counting the number of events whose
arrival directions fall within a circular aperture centered on the
putative source location. The radius of the circular aperture has been
determined by optimization of data taken on the Crab Nebula and is
found to be $0.32\degr$. The analysis of data from an extended region
involves counting the number of events whose arrival directions fall
within a circular aperture which encompasses the emission region plus
twice the width of the angular resolution function to account for
smearing of the image.

In both cases the background is obtained from an equal amount of time
tracking a position offset by 30 minutes in right ascension but with
the same declination as the on source observation.

\begin{table}[tb]
\caption{Flux upper limits for the SNR Cassiopeia A and the Monoceros
SNR - Rosette Nebula region. The result for Cassiopeia A was obtained
by utilizing the method described in section 2.1, while the results
for the Monoceros SNR - Rosette Nebula region were obtained utilizing the
analysis method described in section 2.2.}
\vspace{3mm}
\begin{center}
\begin{tabular}{cccc} \hline
Source     &  $\alpha,\delta$   & aperture  & F(E $>$ 500 GeV) \\
Name       &  (J2000)           & (degrees) & $\times10^{-11} 
                                              {\rm cm}^2 {\rm s}^{-1}$\\ \hline
CAS-A         & 23:23:23,58:48:11  & -         & $<$0.66 \\
Monoceros SNR & 06:35:00,05:21:00  & 1.00      & $<$4.80 \\
- Rosette     & J0635+0533         & 0.32      & $<$1.41 \\                   
\end{tabular}
\end{center}
\end{table}

\section{Results:}
\begin{figwindow}[0,r,
{\epsfig{figure=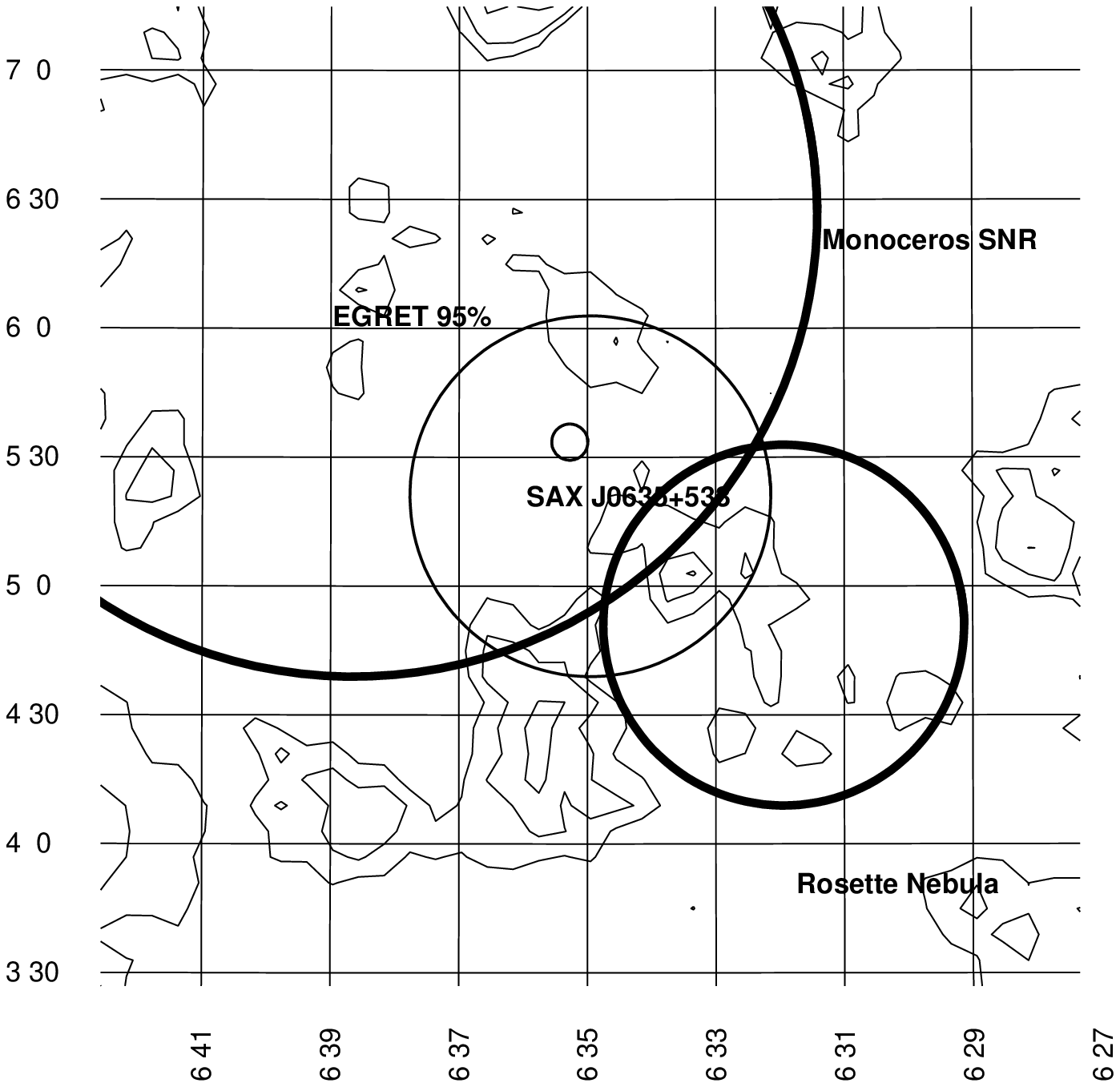, width=9cm}},
{TeV gamma-ray observations of the Monoceros SNR / Rosette Nebula region.
Contours of statistical significance of excess events between
on source observations and corresponding off source observations. The
contours increment by one.}]
\noindent {\bf 4.1 \hspace{2mm} Monoceros SNR - Rosette Nebula Region:}
The Monoceros SNR - Rosette Nebula region was observed for a total of
13.1 hours during the period November 1998 to February 1999.  The
contours depicted in Figure~2 show the statistical significance of the
excess events recorded within the camera's sensitive field of view.
The large thick circle depicts the position and extent of the
Monoceros SNR while the smaller thick circle depicts the position and
extent of the Rosette Nebula (a star forming region). The large thin
circle depicts the EGRET 95\% confidence error circle and the small
circle depicts the position of a hard spectrum X-ray point source,
J0635+0533, recently detected by the Italian-Dutch satellite BeppoSAX
(Kaaret et al. 1999).  No significant excess is detected.  In the
absence of a detection we have calculated flux upper limits based on
two assumptions. First, that the emission detected by EGRET arises
from the X-ray point source J0635+0533. In this case we have
calculated a flux upper limit using the offset point source technique
given above. Second, that the emission detected by EGRET is due to
cosmic rays accelerated by the shock of the Monoceros SNR. In this
case we have calculated a flux upper limit for uniform emission from
the region contained within the EGRET 95\% error circle.  These limits
are given in Table~1.
\end{figwindow}

\noindent {\bf 4.2 \hspace{2mm} Cassiopeia A:}
The SNR, Cassiopeia A, was observed for a total of 6.9 hours during
the period of November 1998 to January 1999. This young remnant
subtends an angle of about $0.08\degr$ which is smaller than the
angular resolution of the photon arrival direction reconstruction.
Hence, we have applied the centered point source method as given
discussed above. No emission is detected and the 99.9\% confidence
level upper limit is given in Table~1.

\section{Discussion:}
\begin{figwindow}[0,r,
{\epsfig{figure=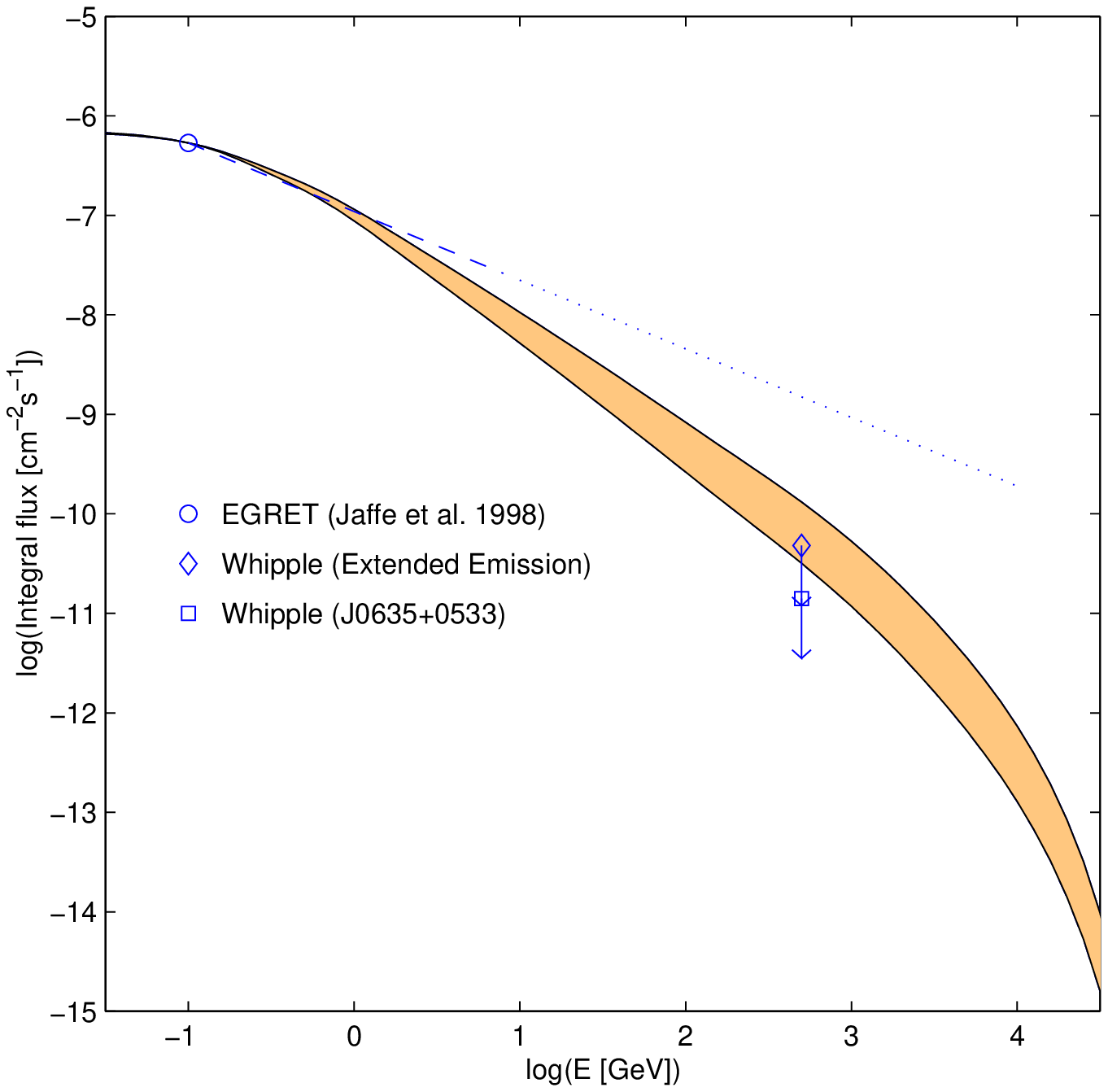, width=9cm}},
{Whipple upper limits for the Monoceros SNR - Rosette Nebula
region shown along with the EGRET integral flux (Jaffe et al. 1998).}]
\noindent {\bf 4.1 \hspace{2mm} The Monoceros SNR - Rosette Nebula Region:}
The Whipple upper limits and EGRET data (Jaffe et al. 1998) for the
Monoceros SNR - Rosette Nebula region are shown in Figure~3. Under the
assumption that the emission detected by EGRET is due to cosmic-ray
interactions we normalize the model of DAV to the EGRET $>100$ MeV
flux. The shaded area depicts a range of the spectral index of the
proton population used in the model (2.1 - 2.3). As shown, the upper
limits reported here require the source spectrum to be steeper than
$E^{-2.2}$ or require a spectral break.
Another possible explanation for the results is that the EGRET flux
is produced not by cosmic-ray interactions but rather by the X-ray
point source J0635+0533 which has been associated with a Be binary
system (Kaaret 1999). In this case, the upper limit reported here lies
two orders of magnitude below the extrapolated EGRET power-law
spectrum and thus requires a spectral break between 10~GeV and
500~GeV.
\end{figwindow}

\noindent {\bf 4.2 \hspace{2mm} Cassiopeia A:}
Non-thermal X-ray emission has recently been observed in the SNR
Cassiopeia~A. This result has been interpreted as synchrotron
radiation by electrons accelerated by the SNR shock wave to energies
of 40 TeV (Allen et al. 1997). These accelerated electrons will also produce
TeV gamma-rays through inverse Compton scattering of soft photons
fields. The results reported here should help to constrain the modeling
of this object.

\subsection*{Acknowledgments:}
We acknowledge the technical assistance of K. Harris and
E. Roache. This research has been supported in part in the U.S. by the
D.O.E and NASA, Enterprise Ireland and PPARC in the UK.

\vspace{1ex}
\begin{center}
{\Large\bf References}
\end{center}
%
Koyama, K., et al. 1995, Nature, 378, 255\\
Allen, G.E., et al. 1997, ApJ, 487, L97\\
Drury, L.O'C., et al. 1994, A\&A, 287, 959\\
Naito, T. \& Takahara, F. 1994, J.Phys. G:Nucl.Part.Phys., 20, 477\\
Esposito, J.A., et al. 1996, ApJ, 461, 820\\
Jaffe, T., et al. 1998, ApJ, 484, L129\\
Buckley, J.H., et al. 1998, A\&A, 329, 639\\
Tanimori, T. et al. 1998, ApJ, 497, L25\\
Pohl, M. 1996, A\&A, 307, 57\\
Mastichiadis, A. \& de~Jager, O.C. 1996, A\&A, 311, L5\\
Cawley, M.F., et al. 1990, Exper. Astr., 1, 173\\
Reynolds, P.T., et al. 1993, ApJ, 404, 206\\
Mohanty, G., et al. 1998, APh, 9, 15\\
Catanese, M., et al. 1998, ApJ, 501, 616\\
Lessard, R.W., et al. 1997, in Towards a Major Atmospheric Cherenkov Detector 
(Kruger Park), 356\\
Kaaret, P., et al. 1999, Astro-ph, 9904249\\

\end{document}